\documentclass{mem}
\usepackage{natbib}\usepackage{txfonts}\usepackage{balance}
\usepackage{graphicx}
\usepackage[a4paper]{hyperref}
\idline{75}{282}
\begin{document}
\def\teff{$T\rm_{eff }$}
\def\kms{$\mathrm {km s}^{-1}$}

\title{
The Metallicity Dependence of the Cepheid Period-Luminosity Relation
}

\subtitle{}

\author{
M. \,Romaniello\inst{1}, F. \,Primas\inst{1}, M. \,Mottini\inst{1},
M. \,Groenewegen\inst{2}, G. \,Bono\inst{3} \and P.~\,Fran\c{c}ois\inst{4}
}

\offprints{M. Romaniello, \texttt{mromanie@eso.org}}

\institute{European Southern Observatory, Karl-Schwarzschild-Strasse 2,
    D--85748 Garching bei M\"unchen, Germany
  \and
    Instituut voor Sterrenkunde, PACS-ICC, Celestijnenlaan 200B,
      B--3001 Leuven, Belgium
  \and
    INAF-Osservatorio Astronomico di Roma, via di Frascati 33, I--00040
      Monte Porzio Catone, Italy
  \and
    Observatoire de Paris-Meudon, GEPI, 61 avenue de l'Observatoire,
      F--75014 Paris, France
}

\authorrunning{Romaniello et al}

\titlerunning{The Cepheid PL relation vs [Fe/H]}

\abstract{
We have assessed the influence of the stellar iron content
on the Cepheid Period-Luminosity ($PL$) relation by relating the $V$
band residuals from the \citet{fre01} $PL$ relation to [Fe/H] for 68
Galactic and Magellanic Cloud Cepheids. The iron abundances were
measured from FEROS and UVES high-resolution and high signal-to-noise
optical spectra. Our data indicate that the stars become fainter as
metallicity increases, until a plateau or turnover point is reached at
about solar metallicity. This behavior appears at odds both with the
PL relation being independent from iron abundance and with Cepheids
becoming monotonically brighter as metallicity increases
\citep[e.g.][]{ken98,sak04}.

}
\maketitle{}

\section{Introduction\label{sec:intro}}
The Cepheid Period-Luminosity ($PL$) relation is undoubtedly a
fundamental tool in determining Galactic and extragalactic
distances. In spite of its paramount importance, to this day we still
lack firm theoretical and empirical assessment on whether or not
chemical composition has any significant influence on it.

We have tackled the problem by measuring the chemical composition of a
total of 68 Cepheids in the Milky Way galaxy and in the Large and
Small Magellanic Cloud from spectra collected with ESO's FEROS and
UVES instruments. The sample and the data reduction and analysis are
thoroughly described in \citet{rom05} and \citet[][see also Mottini et
al, this conference]{mot05}.

\begin{figure}[!h]
\begin{center}
\resizebox{\hsize}{!}{\includegraphics[clip=true]{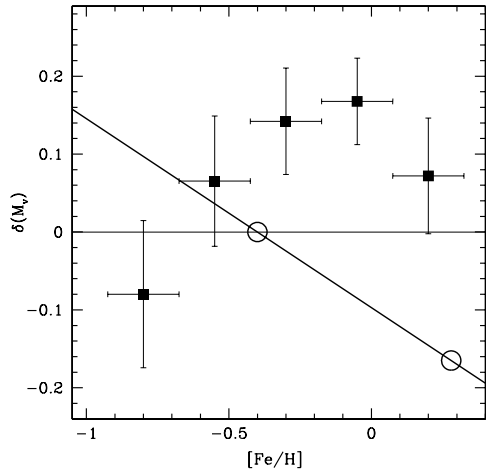}}
\end{center}
\caption{\footnotesize The $V$-band residuals compared to the
\citet{fre01} $PL$ relation are plotted against the iron content
measured from FEROS and UVES spectra. Filled squares represent the
median value in each metallicity bin, with is associated errorbar. The
metallicity dependence as inferred by \citet{ken98} from two Cepheid
fields in M101 (open circles) is shown as a thick line.  }
\label{fig:resz}
\end{figure}

Our main results are summarized in Figure~\ref{fig:resz}, where we
plot the $V$-band residuals compared to the \citet{fre01} $PL$
relation are plotted against the iron content measured from FEROS and
UVES spectra for the 61 stars in our sample with well-determined
distances and photometry and that populate the linear part of the $PL$
relation \citep[$0.4\leq\log(P)\leq1.85$, e.g.][]{bon99}. A positive
$\delta(\mathrm{M}_{\mathrm{V}})$ means fainter than the mean relation.

In computing the $\delta(\mathrm{M}_{\mathrm{V}})$ in
Figure~\ref{fig:resz} we have adopted the periods, distance moduli and
$V$-band photometry of the Galactic Cepheids as listed in Table~3 of
\citet{sto04}. Two of our programme stars, $\zeta~\mathrm{Gem}$ and
$\beta~\mathrm{Dor}$, are not included in that list and for them we
have used the values from \citet[][Table 3]{gro04b}. The periods and
$V$-band photometry for the Magellanic Cloud Cepheids were taken from
\citet{lan94}. The distance modulus of the barycenter of the LMC is
assumed to be 18.50, for consistency with the $PL$ relation of
\citet{fre01}. The SMC is considered 0.44 magnitudes more distant
\citep[e.g.][]{cio00}. Depth and projection effects in the Magellanic
Clouds were corrected for using the position angle and inclination of
each galaxy as determined by \citet[][LMC]{mar01} and
\citet[][SMC]{cal91}.

As it can be seen in Figure~\ref{fig:resz} our data (filled squares)
indicate that Cepheids become fainter as metallicity increases, until
a plateau or turnover point is reached at about solar
metallicity. This trend is at odds both with no dependence of the $PL$
relation upon iron content (thin horizontal line) and with Cepheids
becoming monotonically brighter as metallicity increases
\citep[][thick line]{ken98}.

\begin{acknowledgements}
We warmly thank Giuliana Fiorentino, Lucas Macr\'i, Marcella Marconi
and Ilaria Musella for many useful discussions in the beautiful
setting of the Roman hills.
\end{acknowledgements}

\bibliographystyle{aa}

\end{document}